\documentstyle[12pt]{article}

\newcommand{\be}{\begin{equation}}
\newcommand{\ee}{\end{equation}}
\newcommand{\bea}{\begin{eqnarray}}
\newcommand{\eea}{\end{eqnarray}}
\newcommand{\lb}{\label}

\begin{document}
\begin{titlepage}
\begin{flushright}
 ZU-TH 15/93\\
BUTP-93/15\\
gr-qc/9306029
\end{flushright}
\begin{center}
\vfill
{\large\bf  On time and the quantum-to-classical transition in
Jordan-Brans-Dicke quantum gravity}
\vfill
{\bf Claus Kiefer}\footnote{Address after October, 1st: Institute for
Theoretical Physics, University of Freiburg, Hermann-Herder-Str. 3, D-79104
Freiburg, Germany.}

\vskip 0.4cm
Institute for Theoretical Physics, University of Z\"urich,
Sch\"onberggasse 9,\\CH-8001 Z\"urich, Switzerland
\vskip 0.7cm
{\bf Erik A. Mart\'{\i}nez}\footnote{Address after October, 1st: Theoretical
Physics Institute, Department of Physics, University of Alberta, Edmonton,
Canada T6G 2J1.}
\vskip 0.4cm
Institute for Theoretical Physics, University of Bern, Sidlerstrasse 5,\\
CH-3012 Bern, Switzerland
\end{center}
\vfill
\begin{center}
{\bf Abstract}
\end{center}
\begin{quote}
Any quantum theory of gravity which treats the gravitational constant as a
dynamical variable has to address the issue of superpositions of states
corresponding to different eigenvalues. We show how the unobservability of such
superpositions can be explained through the interaction with other
gravitational degrees of freedom (decoherence). The formal framework is
canonically quantized Jordan-Brans-Dicke theory. We discuss the concepts of
intrinsic time and semiclassical time as well as the possibility of tunneling
into regions corresponding to a negative gravitational constant. We calculate
the reduced density matrix of the Jordan-Brans-Dicke field and show that the
off-diagonal elements can be sufficiently suppressed to be consistent with
experiments. The possible relevance of this mechanism for structure formation
in extended inflation is briefly discussed.

  \end{quote}
\vfill
\begin{center}
June 1993
 \end{center}

\end{titlepage}

It is of fundamental interest to understand the origin of coupling constants in
physics. Some approaches are based on the idea of a {\em dynamical} origin of
these constants within the framework of a more fundamental theory. One
prominent example is the possible effect of wormholes in driving the
cosmological constant to zero \cite{Co}. More generally, wormholes may
dynamically influence any low-energy coupling constant. Since the underlying
framework is that of quantum theory one has to address the issue of {\em
superpositions} of states corresponding to different ``values" of the coupling
constant. Since quantum theory is linear, any a priori selection of a special
state would amount to a selection by hand and therefore prevent the possibility
to understand dynamically the emergence of ``classical" states. It is the
purpose of this letter to provide a dynamical explanation in the particular
example of the gravitational constant. Instead of working with wormholes and
``euclidean" quantum gravity we restrict
ourselves to the simpler framework of Jordan-Brans-Dicke (JBD) theories where
the r\^{o}le of the gravitational constant is played by a scalar field $\phi$
\cite{BD}. Such theories also arise as low energy effective theories from
string theory. We work in the framework of canonical quantization which focuses
on wave functionals depending on the JBD field (apart from the three-metric and
other fields). The issue is then to understand the non-occurrence of
superpositions like
\be \vert\Psi\rangle=\int{\cal D}\phi{\cal F}[\phi]\vert\phi\rangle \label{1}
\ee
which would be in contradiction with the experience of a well-defined
gravitational constant. We will first outline the formal framework of our
investigation and then proceed to understand, within this framework, the
``emergence of the gravitational constant." We note that the effect of
wormholes has also been studied in the context of the JBD theory \cite{GG} with
the conclusion that the wave functional is peaked at infinite JBD parameter
$\omega$, but we will not take into account the effect of any wormholes in the
following.

In its most simple version, the Lagrangian of JBD theories is given by
\be {\cal L}=\sqrt{-g}\left(\phi R-\frac{\omega}{\phi}g^{\mu\nu}\phi,_{\mu}
\phi,_{\nu}\right)+{\cal L}_m, \lb{2} \ee
where $\omega$ is a constant and no mass term or self-interaction for the JBD
field is taken into account. It is straightforward to perform a $3+1$
decomposition to cast the theory into Hamiltonian form. The central equation in
the canonical theory is the Hamiltonian constraint which reads
\bea {\cal H} &=& \frac{1}{\sqrt{h}\phi}\left(\Pi_{ab}\Pi^{ab}-
  \frac{\omega+1}{2\omega+3}(\mbox{Tr}{\bf {\Pi}})^2-
 \frac{\phi\Pi_{\phi}\mbox{Tr}{\bf \Pi}}{2\omega+3}+
 \frac{\phi^2\Pi_{\phi}^2}{2(2\omega+3)}\right)\nonumber\\
& & \ +\sqrt{h}\left(-\phi\, {^{(3)}}\!R + \frac{\omega}{\phi}h^{ab}
  \phi,_a\phi,_b\right) + {\cal H}_m, \lb{3} \eea
where ${\bf \Pi}$ is the geometrodynamical momentum and $\Pi_{\phi}$ is the
momentum canonically conjugate to the JBD field.
We note that the kinetic terms of the JBD field are suppressed in the limit of
large $\omega$ which is the reason why the classical theory goes over to
general relativity for $\omega\to\infty$, provided one chooses the solution
$\phi=constant
=(16\pi G_N)^{-1}$.\footnote{The gravitational constant measured by observation
of slowly moving particles or in time dilation experiments is not $G_N$ but
$(2\omega+4)/(2\omega+3)G_N$ \cite{We}. Since we are interested here only in
the limit of large $\omega$, we will not take this difference into account.}
The coefficients in front of the gravitational momenta are the components of
DeWitt's metric in the space of three-geometries and the JBD field. Due to the
presence of the JBD field it is a $7\times 7$ matrix at each space point. The
components read
\be G_{abcd}=\frac{1}{2\sqrt{h}l^2\phi}\left(h_{ac}h_{bd}
  +h_{ad}h_{bc}-h_{ab}h_{cd}+\frac{h_{ab}h_{cd}}{2\omega+3}\right)
 \lb{4} \ee
and
\be G_{\phi\phi}=\frac{l^2\phi}{\sqrt{h}(2\omega+3)}; \
       G_{ab,\phi}=-\frac{h_{ab}}{\sqrt{h}(2\omega+3)}. \lb{5} \ee
We have introduced here an auxiliary length $l$ so that all components of
DeWitt's metric have the same dimension (note that the momentum canonically
conjugate to $\phi$ has the dimension of an inverse length, whereas the
geometrodynamical momentum has the dimension of an inverse length cube). The
results presented below are independent of $l$.
Viewed as a $7\times 7$ matrix at each space point, the metric can be
diagonalized, with the result
\be G_{AB} = \mbox{diag}\frac{1}{l^2\phi}\left(1,1,\frac{1}{2},\frac{1}{2},
    \frac{1}{2}, \lambda_+(\omega,\phi), \lambda_-(\omega,\phi)\right), \lb{6}
\ee
where
\[ \lambda_{\pm}=\frac{l^4\phi^2 - \omega}{2(2\omega + 3)}
\left(1 \pm \sqrt{1 + \frac{4l^4\phi^2(3 + \omega)}{(l^4\phi^2 - \omega)^2}}
\right) . \]
In the limit of large $\omega$ one finds that $\lambda_+$ and $\lambda_-$
approach the values $-\frac{1}{2}(1-\frac{3}{2\omega})$ and
$l^4\phi^2/2\omega$, respectively. The determinant of (6) is given by
\be \mbox{det}G_{AB} = -\frac{\omega+3}{8l^{10}\phi^5(2\omega+3)^2}.
  \lb{7} \ee
 An interesting point is the behaviour of the signature of this metric. In
general
relativity the signature is hyperbolic with the minus sign arising from the
conformal mode. This minus sign is found from $\lambda_+$ in (6) in the limit
$\omega \to \infty$. But now the signature also depends on the value of the JBD
field and on $\omega$. As can be viewed from (6) and (7), the signature
changes, for positive $\phi$, from hyperbolic to ultrahyperbolic (mixed
signature) if $\omega$ becomes smaller than $-3$. In the case of negative
$\phi$ the overall signature of the metric is changed. This might not seem to
be disturbing, but one must keep in mind that the metric of non gravitational
fields is elliptic so that the hyperbolic nature of the total metric is
destroyed, for negative $\phi$, for all values of $\omega$. While this
behaviour may be irrelevant at the classical level, it has important
consequences for the quantum theory. There the constraint (3) is implemented as
a condition on physically allowed wave functionals, i.e.,
\be \hat{H}\Psi[h_{ab}({\bf x}), \phi({\bf x}),\cdot]=0. \lb{8} \ee
If the signature of DeWitt's metric is hyperbolic, a well-defined initial value
problem can be posed with respect to an ``intrinsic time" variable which is
played by the conformal part of the three-metric \cite{Ze}. This property is
lost in the case of mixed signature. A similar observation has been made before
in the context of scalar fields which are coupled non minimally to gravity
\cite{Ki}. We note that the value $\omega=-1$, which is motivated by string
theory \cite{A} and recently studied models of dilatonic black holes \cite{B},
still lies in the hyperbolic region.
We also note that in more general theories where $\omega$ may depend on $\phi$
the demand for hyperbolicity would lead to a non trivial restriction on the
allowed range of the JBD field. Without such a demand one may have to deal with
the presence of ``superspace Cauchy horizons" like in \cite{Ki} whose
interpretation is not clear.

In the classical theory the particular solution for $\phi$ can be restricted
to lie in the region of positive fields. At first glance, there does not seem
to be a problem at the classical level if $G$ is chosen to be negative which
would correspond to a gravitational repulsion.
 It would, however, lead to a negative ADM energy since the positive energy
theorem is no longer applicable. (A Schwarzschild solution with negative $G$,
for example, would be equivalent to a solution with negative ADM mass.) This is
reflected in the quantum theory in the change of sign in the DeWitt metric.
There the wave functional may tunnel into regions with negative fields. This is
analogous, although different, to the phase space tunneling in the
asymptotically flat context which was discussed by Ashtekar and Horowitz
\cite{AH} and which leads to the possibility of negative energy states in
quantum gravity.

In the following we will first discuss a minisuperspace model (closed Friedmann
universe)
where the gravitational degrees of freedom are the scale factor $a$ and the
homogeneous part $\phi_0$ of the JBD field. Models of this kind have been
discussed in the context of extended inflation e.g. in \cite{QB}.
 The interaction with inhomogeneous degrees of freedom will then be considered
below.
After a field redefinition
\be 2\pi\phi_0=\left(\frac{a}{a_0}\right)^{3/\omega}
  \frac{e^{\chi}}{8G_N} \stackrel{\omega\to\infty}{\to}
  \frac{e^{\chi}}{8G_N} \lb{9} \ee
which diagonalizes the kinetic term,\footnote{Note that the choice of the
variable $\chi$ also prevents the wave function from having support in regions
corresponding to a negative gravitational constant. The general relativistic
value $\phi_0=(16\pi G_N)^{-1}$ corresponds to $\chi=0$.}
the Hamiltonian reads, in the limit of large $\omega$,
\be H_0  =-\frac{G_N\pi_a^2}{3e^{\chi}a} + \frac{2G_N\pi_{\chi}^2}{\omega
e^{\chi}a^3} - \frac{3ae^{\chi}}{4G_N} + \frac{\pi_{\sigma}^2}{2a^3}
+\frac{m^2\sigma^2a^3}{2}. \lb{10} \ee
Here we have added a further massive scalar field $\sigma$ (``inflaton field")
to have a more realistic model. Choosing the Laplace-Beltrami ordering for the
kinetic term, the minisuperspace Wheeler-DeWitt equation reads (with $a\equiv
e^{\alpha}$)
\bea & &  \left(\frac{2G_N}{3}\frac{\partial^2}{\partial\alpha^2}
+ \frac{5G_N}{3}\frac{\partial}{\partial\alpha}- \frac{4G_N}{\omega}
\frac{\partial^2}{\partial\chi^2} - \frac{e^{\chi}}{2}\frac{\partial^2}
 {\partial\sigma^2}\right.\nonumber\\ & & \ \left.
-\frac{3e^{4\alpha}e^{2\chi}}{4G_N} +
\frac{m^2\sigma^2e^{6\alpha}e^{\chi}}{2}\right)\psi_0(\alpha,\chi,\sigma)=0.
\lb{11} \eea
This equation cannot be exactly solved but it can be solved in a
Born-Oppenheimer approximation since $\omega\gg 1$ and the kinetic term of the
$\chi$ field is  suppressed (experimentally one has $\omega > 500$). Since we
do not need the solution for $\psi_0$ explicitly, we will not write it down
here. It is, however, instructive to have a brief look at the minisuperspace
equation in the simpler case where the Friedmann universe is open and the
inflaton field is absent. Instead of (11) we then have (neglecting the linear
factor ordering term)
\be \left(\frac{\omega}{6}\frac{\partial^2}{\partial\alpha^2} -
 \frac{\partial^2}{\partial\chi^2}\right)\psi_0(\alpha,\chi)=0. \lb{12} \ee
This is a wave equation with an effective speed $\sqrt{6/\omega}$ (we consider
$\alpha$ as the time variable). The minisuperspace wave packet therefore does
not spread. Eq.~(12) can of course be easily solved in terms of plane waves,
\be \psi_0(\alpha,\chi)\equiv e^{iS_0(\alpha,\chi)}=e^{ik(\chi-\sqrt
{\frac{6}{\omega}}\alpha)}. \lb{13} \ee
General solutions, for example wave packets, may be constructed by superposing
many plane waves. If one had just one localised packet, this would model the
classical behaviour of this theory and lead to a ``frozen" packet in the
general relativistic limit of infinite $\omega$. As we will see below, however,
the recovery of the time-dependent Schr\"odinger equation for non-gravitational
fields demands the presence of a single WKB state (13). Such a state ascribes
equal probabilities to all values of $\chi$ which would be in sharp
contradiction to the well-defined observed value of the gravitational constant.

How can this apparent contradiction be resolved? The key idea is the
observation that the minisuperspace degrees of freedom of Eq.~(11) are actually
correlated with a huge number of further degrees of freedom, both gravitational
and non-gravitational ones. The relevant object to discuss is therefore not the
minisuperspace wave function but the minisuperspace density matrix which is
obtained by tracing out further, inaccessible, degrees of freedom. It had been
shown before that interferences of different scale factors can become
suppressed through this interaction \cite{Ze2,Ki2}. In fact, due to the
universal coupling of the metric to all degrees of freedom the metric becomes
classical to a high degree of accuracy and we will therefore assume in the
following that the minisuperspace density matrix is already diagonal with
respect to $a$.

How, then, can one understand the emergence of a ``classical" $\phi_0$?
Consider an experiment where the JBD field is determined through the
interaction of two masses. One can of course only measure
-- in full analogy to the measurement of electromagnetic fields \cite{BR} --
 the field average over an appropriate volume $V\approx d^3$, where $d$ is a
typical distance between the masses. The definition of the minisuperspace
density matrix should thus contain a coarse-graining in field space \cite{GH}.
Moreover, the state of the JBD field is quantum correlated with gravitational
fields outside the volume $V$, since $\phi$ is really a cosmological field, in
contrast to, say, the electric field. The gravitational fields in question are
other masses, gravitational waves, and possibly also inhomogeneous degrees of
freedom of $\phi$ itself. These are the degrees of freedom which shall be
traced over and which lead to decoherence.
It should be noted that decoherence for the JBD field is ineffective as far as
non-gravitational degrees of freedom are concerned, since only the metric
couples directly to non-gravitational degrees of freedom.
 In the following discussion we will assume that the coarse-graining has been
already performed, and we will only take the fluctuations in the JBD field
itself into account. It is, however, known that these fluctuations also mimic
the metric degrees of freedom \cite{Ki2}, \cite {HS}.

Technically, the discussion is carried out through an expansion of the full JBD
field into harmonics on the three-sphere  \cite{HH},
\be \phi({\bf x})= \phi_0 + \sum_{nlm}
   f_{nlm}Q^n_{lm}, \lb{14} \ee
and the modes $f_{nlm}\equiv f_n$ are taken into account only up to quadratic
order.
We note that the whole discussion will be carried out in the ``Jordan frame" in
which the metric and the JBD field are not mixed by a field redefinition.
Although it does not matter classically which frame one considers (galaxy
formation in the framework of extended inflation, e.g., is studied easier in
the ``Einstein frame" \cite{ST}), the effect of decoherence is sensitive to the
choice of variables which are summed over. We consider it to be more sensible
if the trace is performed over the original, not the rescaled, variables, since
the unscaled metric is the physical metric which is measured with rods and
clocks.

After a rescaling of the lapse function, which multiplies the constraint (3),
by the JBD field (this will not change the results derived below),  the
Hamiltonian becomes, in the limit of large $\omega$,
\bea H &=& \frac{\pi^2}{2}\left(-\frac{\pi_a^2}{6a}  +
\frac{\phi_0^2\pi_{\phi_0}^2} {\omega a^3}
-\frac{\phi_0\pi_{\phi_0}\pi_a}{\omega a^2} -24a\phi_0^2\right) -
\frac{\pi_a}{4\omega a^2}\sum_n f_n\pi_n \nonumber\\ & &
 +\frac{\phi_0\pi_{\phi_0}}{\omega a^3}\sum_n f_n\pi_n
+\frac{\phi_0^2}{4\omega a^3}\sum_n f_n^2 + \omega a(n^2-1)\sum_n
f_n^2 - 6a\sum_n f_n^2 \nonumber\\
& \equiv & 4\pi^3\phi_0 H_0 +a^{-3}\sum_n H_n,\lb{15} \eea
where $\pi_n$ denotes the momentum canonically conjugate to $f_n$, and we have
not considered any inflaton field.
To solve the Wheeler-DeWitt equation (8) we make the following ansatz for the
wave function
\be \psi\approx \psi_0(a,\phi_0)\prod_{n>0}^N \psi_n(a,\phi_0,f_n), \lb{16} \ee
where $N$ denotes the number of inhomogeneous degrees of freedom.
Assuming that $\psi_0$ is a {\em single} WKB wave function, $\psi_0\approx
e^{iS_0}$, and that the $\psi_n$ depend only adiabatically on $a$ and $\phi_0$,
the wave functions $\psi_n$ obey a Schr\"odinger equation \cite{Ki2}
\be H_n\psi_n\approx i\frac{\partial\psi_n}{\partial t} \lb{17} \ee
with respect to the semiclassical time $t$ which is defined by
\be \frac{\partial}{\partial t}\equiv\nabla S_0\cdot\nabla \lb{18} \ee
($\nabla$ is the gradient in minisuperspace). As already remarked above, the
Schr\"odinger equation can only be recovered if the minisuperspace wave
function is a single WKB state (a possible justification of this fact can be
found in \cite{Ba}).
Neglecting the last term in (15) (since $\omega n^2\gg 6$)
and taking $n\gg1$, the Hamiltonians $H_n$ are given by
\bea H_n & \approx & -\frac{\phi_0^2}{4\omega}\frac{\partial^2}{\partial f_n^2}
 + \frac{ia}{4\omega}\frac{\partial S_0}{\partial a}f_n\frac{\partial}{\partial
f_n}
-\frac{i\phi_0}{\omega}\frac{\partial S_0}{\partial\phi_0}f_n
\frac{\partial}{\partial f_n} \nonumber\\
& & \ +\frac{1}{4\omega}\left(\frac{\partial S_0}{\partial\phi_0}\right)^2f_n^2
       + \omega a^4n^2f_n^2. \lb{19} \eea
Basically, $H_n$ describes a time-dependent oscillator with mass $\omega$.
The terms in (19) containing $S_0$  describe, in the semiclassical limit, the
time variation of the background variables $a$ and $\phi_0$, i.e.
\be \frac{\partial S_0}{\partial a}=-12a^2\phi_0{\cal H}_0 \lb{20} \ee
(where ${\cal H}_0$ is the Hubble parameter), and
\be \frac{\partial S_0}{\partial\phi_0}=\frac{2\omega a^3\dot{\phi_0}}{\phi_0}.
 \lb{21} \ee
In the following we neglect the terms which describe a time variation in the
gravitational constant but keep in mind that these terms would have to be taken
into account if this variation is significant.\footnote{The absolute value of
the ratio of the third to the second term in (19) is given by
$\omega\dot{\phi_0}/6\phi_0
{\cal H}_0$ which is smaller than $\omega/600$ since $\dot{\phi_0}/\phi_0
< 10^{-12}yr^{-1}$ from Viking experiments.}

We now proceed to solve (17) by a Gaussian ansatz and assume that the wave
functions are in their adiabatic ground state. The result is
\be \psi_n=\left(\frac{2\omega a^2 n}{\pi\phi_0}\right)^{1/4}
 \exp\left(-\frac{\omega a^2n}{\phi_0}f_n^2 - \frac{3i{\cal H}_0a^3}{\phi_0}
 f_n^2\right). \lb{22} \ee
We note that the consideration of the terms involving $\dot{\phi_0}$ would
amount to replace $3{\cal H}_0$ by $3{\cal H}_0-2\omega\dot{\phi_0}/\phi_0$.
The minisuperspace density matrix is then obtained by
\be \rho(a,\phi_0,\phi_0')=\prod_{n>0}^N df_n \psi^*_n(a,\phi_0',f_n)
 \psi_n(a,\phi_0,f_n), \lb{23} \ee
where we have already restricted ourselves to the diagonal in $a$- space.
Inserting (22) into (23) one finds
\be \rho(a,\phi_0,\phi_0') \approx \rho_0 \exp\left(-\frac{N^3}{4\phi_0^2}
 (\phi_0-\phi_0')^2 - \frac{27{\cal H}_0^2a^2N}{256\omega^2\phi_0^2}
  (\phi_0-\phi_0')^2 \right), \lb{24} \ee
where $\rho_0$ is the minisuperspace matrix without inhomogeneous degrees of
freedom, and phase factors have not been written out. We emphasize that the
first, dominating, term is independent of the value of $\omega$. As can be seen
from this Gaussian, the influence of a single mode is ineffective since there
would still remain a coherence width in $\phi_0$ which is equal to $\phi_0$
itself. Only a large number of modes can decohere efficiently.

The crucial issue is now the appropriate choice for the maximum number of modes
$N$.
We make the proposition to consider modes only with wavelength bigger than $d$,
the separation of the masses considered above. The remaining wavelengths are
already contained in the coarse-graining of the field over the region $V$
(which we do not discuss in this letter). For $d\approx 1cm$ (laboratory) one
has $N\approx 10^{28}$ while for $d\approx 10^{25}cm$ (supercluster) one has
$N\approx 10^3$. Since
$(\delta\phi_0)^2 = 2\phi_0^2/N^3 \ll \phi_0^2$ both cases are consistent with
the observed ``sharpness" of the gravitational constant, since the remaining
coherence length is much smaller than $\phi_0$ itself.

 We have not tackled in this letter the question of the back reaction of the
inhomogeneous degrees of freedom onto the minisuperspace background. Its
calculation is most conveniently done with the help of the coarse-grained
effective action \cite{HS}, \cite{Mo}. We assume here that this back reaction
turns out to be small at least for sufficiently big values of the scale factor.

We conclude with some brief remarks on the possible relevance of decoherence
for structure formation in the framework of extended inflation \cite{ST}. In
this scenario the inflaton field plays a rather passive role in the sense that
it sits quietly in the false vacuum and thus triggers inflation, while the
quantum fluctuations of the JBD field are the seeds of galaxy formation. The
question then arises at which stage and for which wavelengths these
fluctuations become classical. The idea is to take into account the interaction
of these fluctuations with a ``reservoir" of other gravitational degrees of
freedom which may force the JBD fluctuations to become classical and enable the
formation of galaxies. We hope to return to this issue in a future publication.

\begin{center}
{\bf Acknowledgements}
\end{center}
This work was supported by the Swiss National Science Foundation.

\end{document}